\documentclass[epj]{svjour}
\usepackage{graphics}
\begin{document}
\title{ Phase transitions in a fluid surface model with a deficit angle term}

\author{Hiroshi Koibuchi
}                     
%
%
\institute{Department of Mechanical and Systems Engineering \\
  Ibaraki National College of Technology \\
  Nakane 866, Hitachinaka, Ibaraki 312-8508, Japan }
%
%
\abstract{
Nambu-Goto model is investigated by using the canonical Monte Carlo simulation technique on dynamically triangulated surfaces of spherical topology. We find that the model has four distinct phases; crumpled, branched-polymer, linear, and tubular. The linear phase and the tubular phase appear to be separated by a first-order transition. It is also found that there is no long-range two-dimensional order in the model. In fact, no smooth surface can be seen in the whole region of the curvature modulus $\alpha$, which is the coefficient of the deficit angle term in the Hamiltonian. The bending energy, which is not included in the Hamiltonian, remains large even at sufficiently large $\alpha$ in the tubular phase. On the other hand, the surface is spontaneously compactified into a one-dimensional smooth curve in the linear phase; one of the two degrees of freedom shrinks, and the other degree of freedom remains along the curve. Moreover, we find that the rotational symmetry of the model is spontaneously broken in the tubular phase just as in the same model on the fixed connectivity surfaces. 
}
\authorrunning {H.Koibuchi}
\titlerunning {Phase transition of compartmentalized surface models}
\maketitle
\section{Introduction}\label{intro}
Triangulated surfaces are one of the basic models to investigate the physics of biological membranes and that of strings \cite{WHEATER-JP1994,NELSON-SMMS2004-1,David-TDQGRS-1989,NELSON-SMMS2004-149,Wiese-PTCP2000,Bowick-PREP2001,Gompper-Schick-PTC-1994}. Surface models can exhibit a variety of shapes because of the two-dimensional nature. A well-known model is the one of Helfrich, Polyakov and Kleinert (HPK) \cite{HELFRICH-1973,POLYAKOV-NPB1986,KLEINERT-PLB1986}. Studies have focused on the phase structure of the surface model of HPK \cite{Peliti-Leibler-PRL1985,DavidGuitter-EPL1988,PKN-PRL1988,BKS-PLA2000,BK-PRB2001,GREST-JPIF1991,BOWICK-TRAVESSET-EPJE2001,BCTT-PRL2001}, and it was reported that the model has the smooth phase and the crumpled (or folded) phase on the fixed-connectivity/fluid surfaces \cite{KANTOR-NELSON-PRA1987,KANTOR-SMMS2004,WHEATER-NPB1996,BCFTA-JP96-NPB9697,CATTERALL-NPBSUP1991,AMBJORN-NPB1993,ABGFHHM-PLB1993,BCHHM-NPB9393,KOIB-PLA-20023,KOIB-PLA-2004,KOIB-EPJB-2005,KOIB-EPJB-2006,KD-PRE2002,KOIB-PRE-2004-1,KOIB-PRE-2005-1,KOIB-NPB-2006,KOIB-PRE-2007-1}. 

 A branched-polymer phase and some linear structures, which are not always smooth, can be seen in a fluid model on dynamically triangulated surfaces \cite{KOIB-PRE-2003}. A surface model of Nambu-Goto also has a variety of phases; the crumpled phase, the tubular phase, and the smooth phase, even on the fixed connectivity surfaces of spherical topology when the Hamiltonian includes a deficit angle term \cite{KOIB-PRE-2004-2}, which is an intrinsic curvature energy \cite{BJ-PRD-1993-1994,BEJ-PLB-1993,BIJJ-PLB-1994,FW-PLB-1993}. 

However, little attention has been given to the linear structure (such as a smooth curve) in the surface models. Therefore, we study in this paper the Nambu-Goto model \cite{Nambu} with the deficit angle term on dynamically triangulated surfaces of spherical topology.  

The Nambu-Goto surface model is well known as an ill-defined one whenever the Hamiltonian includes not only no additional term but also the standard bending energy term \cite{ADF}. However, we have already confirmed that the Nambu-Goto surface model changes to a well-defined one on the fixed connectivity surfaces when the deficit angle term is included in the Hamiltonian \cite{KOIB-PRE-2004-2} as well as a certain bending energy term \cite{KOIB-NPB-2006}. Then, it is also interesting to see whether or not such well-definedness remains unaffected on the dynamically triangulated fluid surfaces in the whole range of the curvature coefficient $\alpha (> 0)$ of the deficit angle term. From the standard argument for fluid membranes, we expect that the thermal fluctuations become large on the fluid surfaces. For this reason, we think it is worthwhile to study whether the model is well-defined or not on the fluid surfaces.

The purpose of this study is to understand the phase structure of the surface model of Nambu-Goto with the deficit angle term on dynamically triangulated surfaces of spherical topology. It is also aimed to see whether the model is well defined on the fluid surfaces.

It will be shown in this paper that the model is well-defined and has four distinct phases; the crumpled phase, the branched-polymer phase, the linear phase, and the tubular phase, which consecutively appear as the coefficient $\alpha$ increases from a sufficiently small value ($\alpha\!=\!10$) to a sufficiently large one ($\alpha\!\simeq\!1\!\times\! 10^4$). We will find that the crumpled phase, the branched-polymer phase, and the linear phase are connected by higher-order transitions at relatively small $\alpha$, and that the linear phase and the tubular phase are connected by a first-order transition at relatively large $\alpha$. 

A remarkable result is the appearance of the linear phase, which has not yet been found in the surface models of HPK and in the fixed connectivity Nambu-Goto surface model. In the linear phase of the model in this paper, one of the two dimensions of the surface shrinks around a curve, and the other dimension remains along the curve. Moreover, a spontaneous breakdown of the rotational symmetry is also found in the tubular phase, where the surface spans and extends along a one-dimensional straight line. We have no smooth-phase (or two-dimensional order) in the model even at sufficiently large $\alpha$ in contrast to the same model on the fixed connectivity surfaces \cite{KOIB-PRE-2004-2}. 

\section{Model}\label{model}
The partition function $Z$ of the discrete Nambu-Goto surface model is defined on a triangulated surface and is given by 
\begin{eqnarray}
 \label{Z-Nambu-Goto}
Z(\alpha) = \sum_{\cal T}\int \prod _{i=1}^N dX_i \exp(-S),\quad  \nonumber\\
 S(X,{\cal T})=S_1 + \alpha S_3, \\  
S_1=\sum_{\Delta} A_{\Delta}, \quad S_3=-\sum_i \log (\delta_i/2\pi), \nonumber
\end{eqnarray}
where $\int \prod _{i=1}^N dX_i$ is the $3N$-dimensional integrations, and $\sum_{\cal T}$ denotes the sum over all possible triangulations ${\cal T}$. The Hamiltonian $S$ is given by the linear combination of the area energy term $S_1$ and the deficit angle term $S_3$ such that $S(X,{\cal T})\!=\!S_1 \!+\! \alpha S_3$. $S(X,{\cal T})$ denotes that $S$ depends on the embedding $X$ and the triangulation ${\cal T}$ of the triangulated surface of spherical topology. In the area energy $S_1$ in Eq.(\ref{Z-Nambu-Goto}), $A_{\Delta}$ is the area of the triangle ${\Delta}$. The symbol $\delta_i$ in $S_3$ is the sum of the angles of vertices meeting at the vertex $i$.

The center of the surface is fixed in the integration in Eq.(\ref{Z-Nambu-Goto}) to remove the translational zero mode. $Z(\alpha)$ of Eq.(\ref{Z-Nambu-Goto}) denotes that the model is dependent on the curvature coefficient $\alpha$. It should also be noted that the Hamiltonian $S(X,{\cal T})$ is defined only with intrinsic variables of the surface. 

 The term $S_1$ in Eq.(\ref{Z-Nambu-Goto}) is a discretization of the original Nambu-Goto action defined by $S\!=\!\int d^2x \sqrt{g} $, where $g$ is the determinant of the first fundamental form on the surface $X$ swept out by strings. The surface $X$ is locally understood as a mapping from a two-dimensional parameter space into ${\bf R}^3$. The mapping $X$ is not always injective, because the surface is allowed to self-intersect in string models. Therefore, the model can be called the phantom surface model as long as it is considered as a model for membranes in ${\bf R}^3$. Note also that the Polyakov string model can be obtained from the original Nambu-Goto model \cite{Polyakov-contempV3-1987}.  
 
The deficit angle term $S_3$ is closely related to the integration measure $dX_i$ \cite{David-NP,BKKM}. We have to remind ourselves of that $dX_i$ can be replaced by the weighted measure $dX_iq_i^\alpha $, where $q_i$ is the co-ordination number of the vertex $i$, and $\alpha $ is considered to be $\alpha\!=\!3/2 $. Considering that $q_i$ is a volume weight of the vertex $i$, we assume $\alpha $ as an arbitrary number. Moreover, the co-ordination number $q_i$ can be replaced  by the vertex angle $\delta_i$ such that 
\begin{equation}
 \Pi_i dX_i q_i^\alpha \to \Pi_i dX_i \exp(\alpha \sum_i \log \delta_i). \nonumber
\end{equation}
The constant term $ \sum_i \log 2\pi$ is included to normalize $S_3$ in Eq.(\ref{Z-Nambu-Goto}) so that $S_3\!=\!0$ when $\delta_i\!=\!2\pi$ at the vertices. Thus, we have the expression $S_3$ in Eq.(\ref{Z-Nambu-Goto}). Note that $S_3\!=\!0$ is satisfied not only on the flat surface but also on the cylindrical (or tubular) surfaces. 

The unit of physical quantities is as follows: The length unit $a$ in the model can arbitrarily be fixed because of the scale invariant property of the partition function. As a consequence, the string tension coefficient $\lambda$ in $S\!=\!\lambda S_1\!+\!\alpha S_3$ can be fixed to $\lambda\!=\!1$, because $\lambda$ has the unit of $kT/a^2$, where $k$ is the Boltzmann constant and $T$ is the temperature. The coefficient $\alpha$ has the unit of $kT$.

\section{Monte Carlo technique}\label{MC-Techniques}
The icosahedron is used to construct the triangulated lattices for the starting configuration of the Monte Carlo (MC) simulations. The edges of the icosahedron are divided into $\ell$ pieces of uniform length, then we have a triangulated lattice of size $N\!=\!10\ell^2+2$. 

The canonical Metropolis technique is used to update $X$, which is the three-dimensional vertex position of the triangulated surface. The position $X_i$ is shifted such that  $X_i^\prime\!=\!X_i\!+\!{\mit \Delta} X_i$, where ${\mit \Delta} X_i$ is randomly chosen in a small sphere. The new position $X_i^\prime$ is accepted with the probability ${\rm Min}[1,\exp\left(-{\mit \Delta}S\right) ]$, where ${\mit \Delta}S\!=\!S({\rm new})\!-\!S({\rm old})$. The radius of the small sphere for ${\mit \Delta} X_i$ is chosen at the beginning of the simulations to maintain 35 $\sim$ 65 $\%$ acceptance rate; almost all MC simulations are performed on about 50 $\%$ acceptance rate.

The variable ${\cal T}$ is summed over by the bond flip technique, which destroys the uniform lattice structure of the starting configurations constructed from the icosahedron. The vertices freely diffuse over the surface and, therefore, such surfaces are called the fluid random surface. A sequential number labeling the vertices and another sequential number labeling the bonds become at random, because the bond flips change the pairing between a bond and the corresponding two vertices. By using the sequential number of bonds, a bond is randomly selected to be flipped, and the flip is accepted with the probability ${\rm Min}[1,\exp\left(-{\mit \Delta}S\right) ]$; this is one update of ${\cal T}$. $N$ updates of $X$ and $N$ updates of ${\cal T}$ are consecutively performed and make one MCS (Monte Carlo sweep). The acceptance rate $r_{\cal T}$ of the bond flip is not fixed a priori. We have $r_{\cal T}\!=\!0.65$ at $\alpha\!=\! 10$ in the crumpled phase, and it increases with increasing $\alpha$ and is almost independent of $\alpha$  at $\alpha\!\geq\! 30$, where $r_{\cal T}\!=\!0.53\sim0.55$. 

The lower bound $10^{-6}A_0$ is imposed on the area of triangles in the updates of $X$ and ${\cal T}$, where $A_0$ is the mean area of the triangles computed at every 500 MCS, and $A_0$ remains constant due to the relation $S_1/N\!=\!1.5$. We remark that the areas are almost free from the lower bound, because the areas of almost all triangles are larger than $10^{-6}A_0$ throughout the MC simulations. No constraint is imposed on the bond length. A sequence of random numbers called Mersene-Twister \cite{Matsumoto-Nishimura-1998} is used to update $X$ and ${\cal T}$ in the simulations.

\section{Results}\label{Results}

\begin{figure}[hbt]
\unitlength 0.1in
\begin{picture}( 0,0)(  10,10)
\put(16,41.5){\makebox(0,0){(a) $\alpha\!=\!10(N\!=\!1442)$ }}%
\put(33,41.5){\makebox(0,0){(b) $\alpha\!=\!50(N\!=\!1442)$ }}%
\put(16.5,8.5){\makebox(0,0){(c) $\alpha\!=\!2000(N\!=\!1442)$ }}%
\put(33.5,8.5){\makebox(0,0){(d) $\alpha\!=\!3000(N\!=\!1442)$ }}%
\end{picture}%
\vspace{0.5cm}
\resizebox{0.49\textwidth}{!}{%
\includegraphics{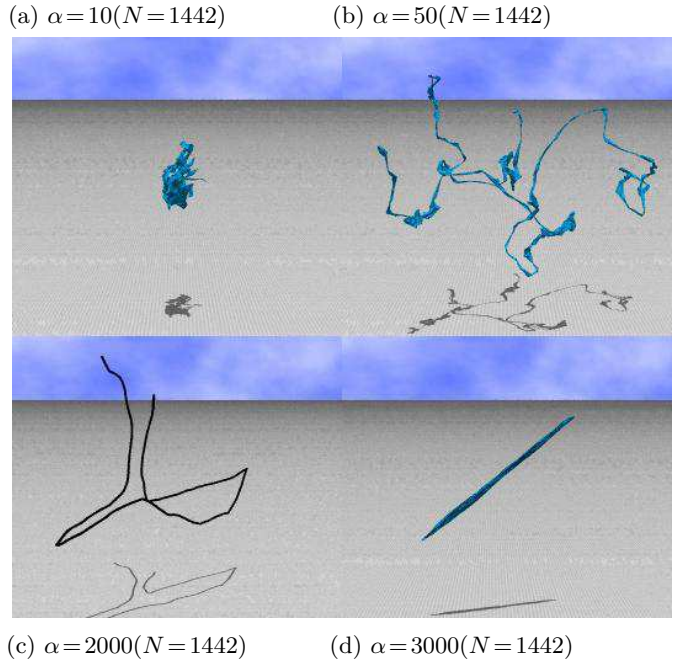}
}
\caption{Snapshots of the $N\!=\!1442$ surfaces obtained at (a) $\alpha\!=\!10$ (crumpled phase),  (b) $\alpha\!=\!50$ (branched-polymer phase), (c) $\alpha\!=\!2000$ (linear phase), and (d) $\alpha\!=\!3000$ (tubular phase). Figures (a),(b), and (d) were drawn in the same scale, which is different from that of (c). The thickness of the surface in (c) is drawn many times larger than the original one, which is too thin to draw.} 
\label{fig-1}
\end{figure}

\begin{figure}[hbt]
\unitlength 0.1in
\begin{picture}( 0,0)(  10,10)
\put(16,42.0){\makebox(0,0){(a) $\alpha\!=\!10(N\!=\!2892)$ }}%
\put(33,42.0){\makebox(0,0){(b) $\alpha\!=\!50(N\!=\!2892)$ }}%
\put(16.5,8.5){\makebox(0,0){(c) $\alpha\!=\!5000(N\!=\!2892)$ }}%
\put(33.5,8.5){\makebox(0,0){(d) $\alpha\!=\!6000(N\!=\!2892)$ }}%
\end{picture}%
\vspace{0.5cm}
\resizebox{0.49\textwidth}{!}{%
\includegraphics{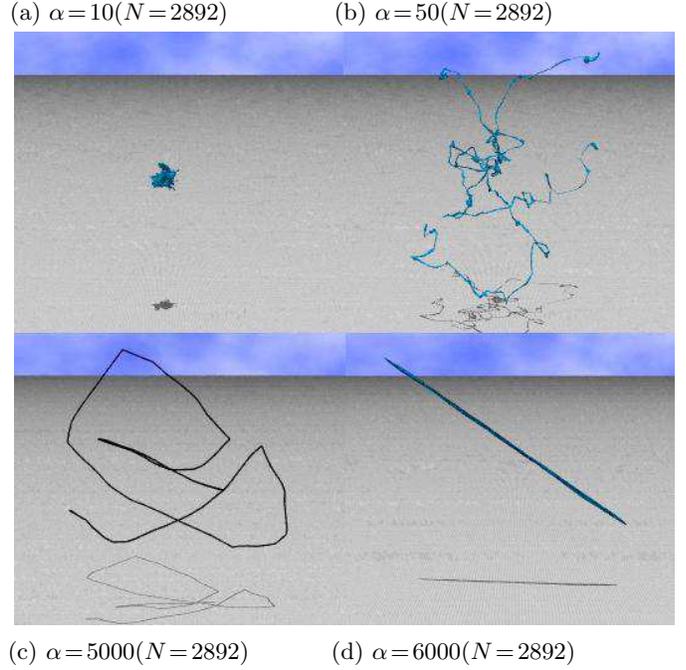}
}
\caption{Snapshots of the $N\!=\!2892$ surfaces obtained at (a) $\alpha\!=\!10$ (crumpled phase),  (b) $\alpha\!=\!50$ (branched-polymer phase), (c) $\alpha\!=\!5000$ (linear phase), and (d) $\alpha\!=\!6000$ (tubular phase). Figures (a),(b), and (d) were drawn in the same scale, which is different from that of (c). The thickness of the surface in (c) is drawn many times larger than the original one, which is too thin to draw. } 
\label{fig-2}
\end{figure}
Snapshots of the $N\!=\!1442$ surfaces are shown in Figs. \ref{fig-1}(a), \ref{fig-1}(b), \ref{fig-1}(c), and \ref{fig-1}(d), where the surfaces were respectively obtained at $\alpha\!=\!10$ (crumpled phase), $\alpha\!=\!50$ (branched-polymer phase), $\alpha\!=\!2000$ (linear phase), and $\alpha\!=\!3000$ (tubular phase). Figures \ref{fig-1}(a), \ref{fig-1}(b), and \ref{fig-1}(d) were drawn in the same scale, which is different from that of Fig.\ref{fig-1}(c). The thickness of the linear surface in Fig.\ref{fig-1}(c) is drawn many times larger than the original one, which is too thin to draw. The reason why we call the surface in Fig.\ref{fig-1}(b) as the {\it branched-polymer} surface is because of its shape. Although the surface in Fig.\ref{fig-1}(b) appears to be almost linear, we can see that some parts of the surface are branched. 

Figures \ref{fig-2}(a), \ref{fig-2}(b), \ref{fig-2}(c), and \ref{fig-2}(d) show snapshots of the $N\!=\!2892$ surface, which were obtained respectively at $\alpha\!=\!10$ (crumpled phase), $\alpha\!=\!50$ (branched-polymer phase), $\alpha\!=\!5000$ (linear phase), and $\alpha\!=\!6000$ (tubular phase). Figures \ref{fig-2}(a), \ref{fig-2}(b), and \ref{fig-2}(d) were drawn in the same scale,  while the scale of Fig.\ref{fig-2}(c) is different from that of the other three figures.  The thickness of the linear surface in Fig.\ref{fig-2}(c) is drawn many times larger than the original one, which is too thin to draw. 

The mean square size $X^2$ is defined by
\begin{equation}
\label{Mean-Square-Size}
X^2= {1\over N} \sum _i \left( X_i-\bar X \right)^2,\quad \bar X = {1\over N} \sum_i X_i,
\end{equation}
where $\bar X$ is the center of the surface. 

\begin{figure}[htb]
\resizebox{0.49\textwidth}{!}{%
\includegraphics{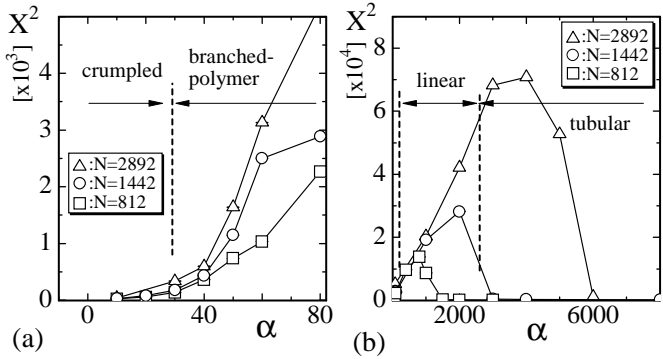}
}
\caption{(a) $X^2$ obtained at relatively small $\alpha$ close to the boundary between the crumpled phase and the branched-polymer phase, and (b) $X^2$ obtained at relatively large $\alpha$ close to the boundary between the linear phase and the tubular phase. The dashed lines drawn vertically in both of the figures represent the phase boundaries of the $N\!=\!1442$ surface. }
\label{fig-3}
\end{figure}
Figure \ref{fig-3}(a) shows $X^2$ obtained at relatively small $\alpha$ close to the phase boundary between the crumpled phase and the branched-polymer phase, where the surface size is $N\!=\!812$, $N\!=\!1442$ and $N\!=\!2892$. In Fig.\ref{fig-3}(b), we plot $X^2$ obtained at relatively large $\alpha$ close to the phase boundary between the linear phase and the tubular phase, where three different sizes;  $N\!=\!812$, $N\!=\!1442$ and  $N\!=\!2882$,  were also assumed. The curves in Fig.\ref{fig-3}(a) show that the surface size continuously increases with increasing $\alpha$. On the contrary, $X^2$ shown in Fig.\ref{fig-3}(b) abruptly changes and then, it clearly indicates a discontinuous transition between the linear phase and the tubular phase. 

The dashed lines drawn vertically in both of the figures represent the phase boundaries of the $N\!=\!1442$ surface. The position of the phase boundary in Fig.\ref{fig-3}(b) is immediately found between the linear phase and the tubular phase, because $X^2$ discontinuously changes at the boundary. On the contrary, the phase boundary between the crumpled phase and the branched-polymer phase in Fig.\ref{fig-3}(a) is unclear. Therefore, the position of it was roughly determined by viewing the snapshots of surfaces at the boundary region of $\alpha$.

The branched-polymer phase and the linear phase are also ambiguously separated; they are expected to connect smoothly to each other, because $X^2$ continuously changes at the corresponding region of $\alpha$ as can be seen in Figs.\ref{fig-3}(a) and \ref{fig-3}(b). The position of the phase boundary of the $N\!=\!1442$ surface is at $\alpha\!=\!80\sim 100$, which was also expected by viewing the snapshots of surfaces.  

The convergence speed of MC simulations is very low in the linear phase close to the tubular phase. About $6\times 10^9 \sim 8\times 10^9$ MCS were done for the thermalization at $\alpha\!=\!3000\sim 5000$ on the $N\!=\!2892$ surface. One reason of such low convergence speed in the linear phase seems due to the large phase-space volume in ${\bf R}^3$. The crumpled surfaces in the crumpled phase occupy a relatively small region in ${\bf R}^3$, while long, thin, and string-like surfaces in the linear phase can extend to a large space in ${\bf R}^3$. Then, the canonical and local update procedure of $X$ is very time-consuming for such string-like surfaces to have the equilibrium configurations as long as the surface obeys the Hamiltonian of short ranged interactions.  

\begin{figure}[htb]
\resizebox{0.49\textwidth}{!}{%
\includegraphics{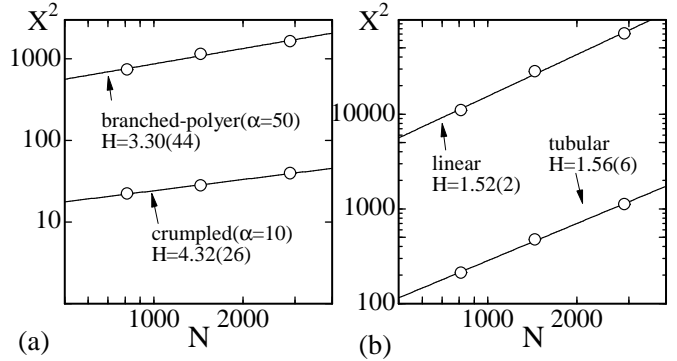}
}
\caption{(a) Log-log plots of $X^2$ against $N$ obtained at $\alpha\!=\!10$ and $\alpha\!=\!50$, and (b) log-log plots of $X^2$ against $N$ obtained at $\alpha$ where $X^2$ is at its maximum in the linear phase and at  $\alpha$ in the tubular phase close to the linear phase. }
\label{fig-4}
\end{figure}
We comment on the Hausdorff dimension $H$, which is defined by the relation
\begin{equation}
\label{X2-scale}
X^2 \propto N^{2/H}
\end{equation}
at sufficiently large $N$. Figures \ref{fig-4}(a) and \ref{fig-4}(b) are log-log plots of $X^2$ against $N$. The data $X^2$ in Fig.\ref{fig-4}(a) were obtained at $\alpha\!=\!10$ in the crumpled phase and at $\alpha\!=\!50$ in the branched-polymer phase, and the data $X^2$ in Fig.\ref{fig-4}(b) were obtained at $\alpha$ where $X^2$ is at its maximum in the linear phase and at $\alpha$ in the tubular phase close to the linear phase. Thus, we have
\begin{eqnarray}
\label{Hausdorff-results}
H_{\alpha\!=\!10} = 4.32\pm 0.26, \quad H_{\alpha\!=\!50} =3.30\pm 0.44, \nonumber \\ 
H_{\rm lin }=1.52\pm 0.02, \quad H_{\rm tub}=1.56\pm 0.06. 
\end{eqnarray}

It is quite natural that the result $H_{\alpha\!=\!10} \!=\! 4.32(26)$ is larger than the physical bound $H\!=\!3$ in the crumpled phase, because the surfaces are allowed to self-intersect and completely crumpled. The result $H_{\alpha\!=\!50} \!=\!3.30(44)$ is larger than the value $H\!=\!2$, which is specific to the branched-polymer surfaces. This indicates that the branched-polymer phase of the model is not exactly the branched-polymer phase. However, we call the surface obtained at that region of $\alpha$ as the branched-polymer surface because of its surface-shape such as shown in Figs.\ref{fig-1}(b) and \ref{fig-2}(b), as stated above.
 
We find that $H_{\rm lin }$, which was obtained in the linear phase close to the tubular phase, is almost identical to $H_{\rm tub}$, which was obtained in the tubular phase close to the linear phase. We understand that both results $H_{\rm lin }\!=\!1.52(2)$ and $H_{\rm tub}\!=\!1.56(6)$ are larger than the value $H\!=\!1$ of a one-dimensional straight line of constant density of vertices and less than the value $H\!=\!2$ of the branched polymer surfaces, and moreover,  $H_{\rm lin }$ and $H_{\rm tub}$ are slightly larger than $H\!=\!1.22(33)$ in the tubular phase of the fixed connectivity surface model \cite{KOIB-PRE-2004-2}. 

The size of the surface can also be measured by the maximum linear size $D$, which can be approximated as follows: Firstly, find the vertex $I$ that has the maximum distance from $\bar X$ the center of surface, and secondly, find the vertex $J$ that has the maximum distance $D$ from the vertex $I$. Then, we have $D=\sqrt{(X_I-X_J)^2}$. Note that $D$ is identical to the diameter when the surface is a sphere. 

\begin{figure}[htb]
\resizebox{0.49\textwidth}{!}{%
\includegraphics{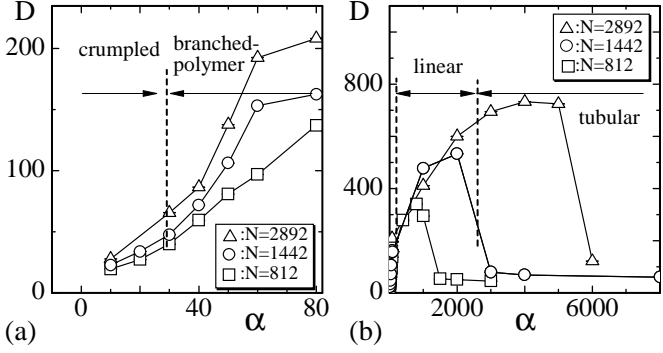}
}
\caption{The maximum linear size $D$ obtained at (a) relatively small $\alpha$ close to the boundary between the crumpled phase and the branched-polymer phase and at (b) relatively large $\alpha$ close to the boundary between the linear phase and the tubular phase. The dashed lines drawn vertically in both of the figures represent the phase boundaries of the $N\!=\!1442$ surface.}
\label{fig-5}
\end{figure}
Figures \ref{fig-5}(a) and \ref{fig-5}(b) show $D$ of the surfaces at $\alpha$ close to the phase boundaries. The size $D$ changes discontinuously at the transition point in Fig.\ref{fig-5}(b). We can also understand from Fig.\ref{fig-5}(b) that the string-like surfaces curve and entangle themselves to occupy a spherical region in ${\bf R}^3$ in the linear phase, since $D(2892)/D(1442)\!\simeq\!1.37({\rm lin})$ is greater than $1.26 \!\simeq \!(2892/1442)^{1/3}$ and less than $2.01\!\simeq\!2892/1442$. The value $1.26$ for the ratio $D(2892)/D(1442)$ is given by assuming that the string-like surface, whose length is proportional to $N$, forms a spherical region in ${\bf R}^3$. If the region has a constant density of vertices, then the diameter of the region is proportional to $N^{1/3}$. In this case, the corresponding Hausdorff dimension is $H\!=\!3$. On the other hand, the value $2.01$ for the ratio $D(2892)/D(1442)$ is given by assuming that the string-like surface spans a one-dimensional straight line of constant density. The corresponding Hausdorff dimension is $H\!=\!1$ in this case.

We also find from Fig.\ref{fig-5}(b) that $D(2892)/D(1442)\!\simeq\!1.53({\rm tub}) $ in the tubular phase close to the linear phase. The value $1.53({\rm tub})$ is greater than $1.42\!\simeq\!\sqrt{2892/1442}$ and less than $2.01\!\simeq\!2892/1442$. The value $1.42$ ($2.01$) for the ratio $D(2892)/D(1442)$ is given by assuming that the surface forms a tubular surface, whose density of vertices is proportional to $N^{1/2}$ ($N^0$) per unit-length, and therefore, the corresponding Hausdorff dimension is $H\!=\!2$ ($H\!=\!1$).  

\begin{figure}[htb]
\resizebox{0.49\textwidth}{!}{%
\includegraphics{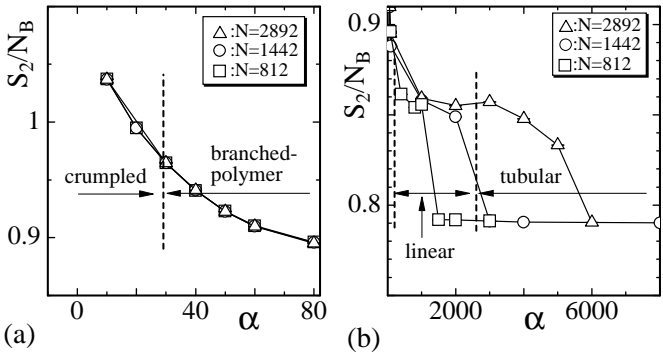}
}
\caption{The bending energy $S_2/N_B$ obtained at (a) relatively small $\alpha$ and at (b) relatively large $\alpha$, where $N_B$ is the total number of bonds. The bending energy $S_2$ is not included in the Hamiltonian. }
\label{fig-6}
\end{figure}
The bending energy $S_2/N_B$ is plotted in Figs.\ref{fig-6}(a) and \ref{fig-6}(b), where $N_B(\!=\!3N\!-\!6)$ is the total number of bonds, and $S_2$ is given by $S_2\!=\!\sum_{ij}(1-{\bf n}_i\cdot{\bf n}_j)$, where $\sum_{ij}$ denotes the summation over unit normal vectors ${\bf n}_i$ and ${\bf n}_j$ of the triangles $i$ and $j$ sharing a common bond. Although $S_2$ is not included in the Hamiltonian, it can reflect how smooth the surface is. We find from Fig.\ref{fig-6}(a) that no discontinuous change can be seen in $S_2/N_B$. On the contrary, $S_2/N_B$ in Fig.\ref{fig-6}(b) appears to change discontinuously at the transition point between the linear phase and the tubular phase. This indicates that the transition is of first order, although the value of $S_2/N_B$ is relatively large even in the tubular phase. The large value of $S_2/N_B$ implies that the surface is not smooth even at $\alpha\!=\!8000$ on the $N\!=\!1442$ surface. 

It is interesting to see whether the smooth surface appears at sufficiently large $\alpha$. Therefore, we performed MC simulations at $\alpha\!=\!1\times 10^4$, $\alpha\!=\!2\times 10^4$, $\alpha\!=\!3\times 10^4$, $\alpha\!=\!5\times 10^4$, $\alpha\!=\! 1\times 10^5$ on the $N\!=\!1442$ surface. As a consequence, we found that the obtained $S_2/N_B$ are almost identical to those in the tubular phase in Fig.\ref{fig-6}(b). This implies that the smooth surface can be seen only at $\alpha\!\to\! \infty$. Thus, we conclude that no smooth phase can be seen in the model on fluid surfaces in contrast to the same model on the fixed connectivity surfaces, where the smooth phase can be seen at sufficiently large but finite $\alpha$. 

\begin{figure}[htb]
\resizebox{0.49\textwidth}{!}{%
\includegraphics{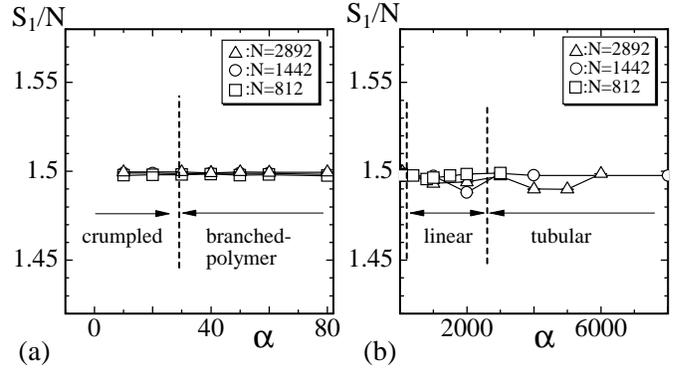}
}
\caption{The area energy $S_1/N$ obtained at (a) relatively small $\alpha$ and at (b) relatively large $\alpha$.  }
\label{fig-7}
\end{figure}
In order to see that the model is well-defined, we plot $S_1/N$ versus $\alpha$ in Figs.\ref{fig-7}(a) and \ref{fig-7}(b). We find from the figures that $S_1/N$ satisfies the expected relation $S_1/N\!=\!3(N-1)/2N\!\simeq\!3/2$. If the relation was violated, then the equilibrium statistical condition should not be expected in the model. In fact, we are unable to see the relation $S_1/N\!=\!3/2$ in an ill-defined model such as the one defined only by the term $S_1$ in Eq.(\ref{Z-Nambu-Goto}). Thus, the results shown Figs.\ref{fig-7}(a) and \ref{fig-7}(b) indicate that the model in this paper is well-defined.   

\begin{figure}[htb]
\resizebox{0.49\textwidth}{!}{%
\includegraphics{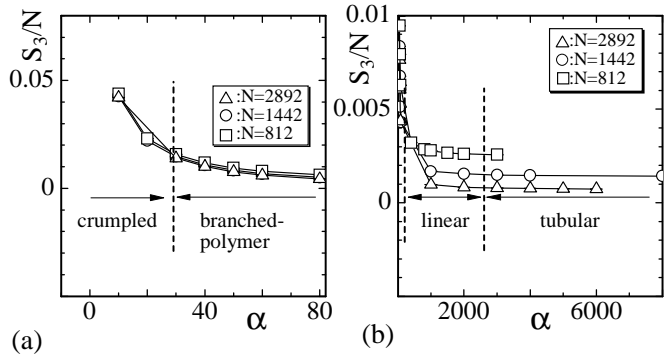}
}
\caption{The deficit angle term $S_3/N$ obtained at (a) relatively small $\alpha$ and at (b) relatively large $\alpha$. }
\label{fig-8}
\end{figure}
The value of the deficit angle term per vertex $S_3/N$ is plotted in Figs.\ref{fig-8}(a) and \ref{fig-8}(b). If the transition is of first order, one can see a discontinuity in $S_3/N$. However, $S_3/N$ appears to change smoothly against $\alpha$ at the phase boundaries. We hardly see a discontinuous change in $S_3/N$ even at the discontinuous transition point in Fig.\ref{fig-8}(b). One reason of this seems come from the fact that the coefficient $\alpha$ is very large while $S_3$ is very small at the transition point. In fact, the Hamiltonian is able to have a finite jump even when $S_3$ has a small jump because of the large value of $\alpha$. If we have many simulation data at the phase boundary between the linear phase and the tubular phase, a discontinuity can be seen in $S_3/N$. However, we must note that the possibility of continuous transition is not completely eliminated. 

\begin{figure}[htb]
\resizebox{0.49\textwidth}{!}{%
\includegraphics{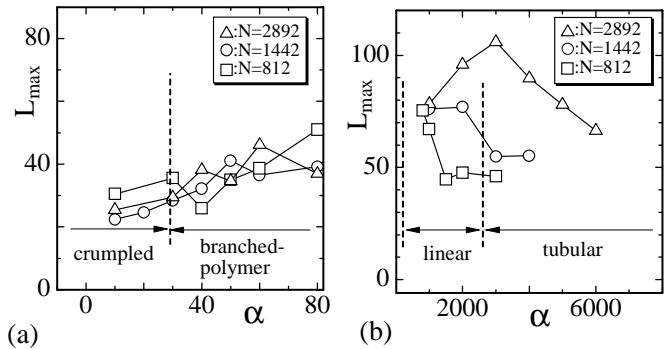}
}
\caption{The maximum bond length $L_{\rm max}$ obtained at (a) relatively small $\alpha$, and at (b) relatively large $\alpha$. $L_{\rm max}$ were obtained in the final $2\times 10^7$ MCS at each $\alpha$.}
\label{fig-9}
\end{figure}
The maximum bond length $L_{\rm max}$ is expected to change depending on $\alpha$. In contrast to the standard Gaussian bond potential, the area energy $S_1$ gives a constraint only on the area of triangles and no constraint on the bond length in the Nambu-Goto model. Consequently, long and thin triangles appear in the equilibrium configurations and form such linear surfaces. Therefore, we have to check that $L_{\rm max} \!<\! D$; the maximum bond length is less than the linear size $D$ except in the crumpled phase, where $D$ can be comparable to $L_{\rm max}$ because of the crumpled nature. In order to see this, we plot $L_{\rm max}$ versus $\alpha$ in Figs.\ref{fig-9}(a) and \ref{fig-9}(b). These $L_{\rm max}$ were obtained in the final $2\times 10^7$ MCS at each $\alpha$. We find from Figs.\ref{fig-9} and \ref{fig-5} that $L_{\rm max}$ is comparable to $D$ in the crumpled phase as expected, and that $L_{\rm max}$ is quite larger than $D$ in the linear phase. Moreover, $L_{\rm max}$ is still larger than $D$ in the tubular phase. In fact, $D\!=\!79$, $L_{\rm max}\!=\!55$ at $\alpha\!=\!3000$ on the $N\!=\!1442$ surface, and $D\!=\!121$, $L_{\rm max}\!=\!66$ at $\alpha\!=\!6000$ on the $N\!=\!2892$ surface. 

\begin{figure}[htb]
\resizebox{0.49\textwidth}{!}{%
\includegraphics{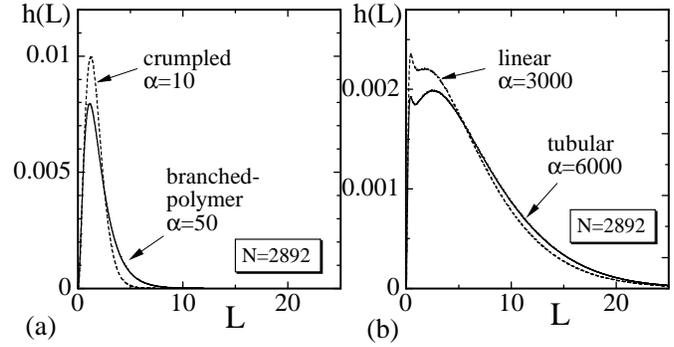}
}
\caption{(a) The histogram $h(L)$ of the bond length $L$ of triangles at the phase boundary between the crumpled phase and  the branched-polymer phase, and (a) those at the phase boundary between the linear phase and the tubular phase. The samples of the bond length $L$  were obtained at every $1000$ MCS in the final $2\times 10^7$ MCS. }
\label{fig-10}
\end{figure}
Not only the surface shape but also the surface structure can be influenced by the phase transitions and changes depending on the value of $\alpha$. Figures \ref{fig-10}(a) and \ref{fig-10}(b) show the distribution (or histogram) $h(L)$ of the bond length $L$ obtained at $\alpha$ close to the phase boundaries. The histogram $h(L)$ was obtained during the final $2\times 10^7$ MCS at each $\alpha$, and $h(L)$ is normalized such that the graph $h(L)$ and the horizontal axis enclose a constant area. We find from Fig.\ref{fig-10}(a) that $h(L)$ at $\alpha\!=\!10$ (the crumpled phase) is different from $h(L)$ at $\alpha\!=\!50$ (the branched-polymer phase). This fact indicates that the lattice structure in the crumpled phase is different from that of the branched-polymer phase, although $h(L)$ is expected to change smoothly from one phase to the other phase.

It is also found from Fig.\ref{fig-10}(b) that the histogram $h(L)$ in the linear phase ($\alpha\!=\!3000$) is different from the one in the tubular phase ($\alpha\!=\!6000$). We can expect that the shape of $h(L)$  discontinuously changes because of the discontinuous nature of the transition, in fact, it is obvious that $h(L)$ at one phase boundary in Fig.\ref{fig-10}(a) is quite different from $h(L)$ at another phase boundary in Fig.\ref{fig-10}(b). However, $h(L)$ shown in Fig.\ref{fig-10}(b) appears to vary smoothly from one phase to the other. The reason of this seems because the surfaces are composed of oblong triangles both in the linear phase and in the tubular phase. We must recall that $h(L)$ in the fixed connectivity model discontinuously changes at the transition point between the tubular phase and the smooth phase \cite{KOIB-PRE-2004-2}; the reason of the discontinuous change of $h(L)$ is because the triangles in the tubular phase are oblong while those in the smooth phase are almost regular. 

We remark that the histogram $h(A)$ of the area $A$ of triangles is not influenced by the phase transitions in contrast to $h(L)$, just as in the fixed connectivity model in \cite{KOIB-PRE-2004-2}. In fact, we checked that no difference can be seen in the histograms $h(A)$ obtained in the crumpled phase, in the branched-polymer phase, in the linear phase, and in the tubular phase on the surfaces of size $N\!=\!812$, $N\!=\!1442$, and $N\!=\!2892$. 

\section{Summary and conclusion}\label{Conclusions} 
A surface model defined by the Nambu-Goto Hamiltonian is investigated by MC simulations on dynamically triangulated surfaces of size up to $N\!=\!2892$. The Hamiltonian includes a deficit angle term. The Nambu-Goto surface model is well known as an ill-defined one if the Hamiltonian includes not only no additional term but also the standard bending energy term. It was also reported that the deficit angle term makes the Nambu-Goto surface model well-defined on the fixed connectivity surfaces of spherical topology, and the phase structure of the model was clarified \cite{KOIB-PRE-2004-2}.

In this paper, we aimed at showing whether the model is well defined or not on the fluid surfaces and how the phase structure changes when the surface changes from the fixed connectivity one to the fluid one. The fluid surface model in this paper has a single parameter $\alpha$ just like in the fixed connectivity model in \cite{KOIB-PRE-2004-2}. The only difference between the model in \cite{KOIB-PRE-2004-2} and the one in this paper is that the model in \cite{KOIB-PRE-2004-2} is defined on the fixed connectivity surface while the model in this paper is defined on the fluid surface.

We found that the fluid model in this paper is well defined and has four distinct phases; the crumpled phase, the branched-polymer phase, the linear phase, and the tubular phase. The first three are smoothly connected, and the last two are connected by a first-order phase transition. We should note that the possibility of higher-order transition between the linear phase and the tubular phase was not completely eliminated. There is no smooth phase in the whole range of $\alpha (>0)$. 

One remarkable result is that the model has the linear phase, where the surface shrinks to such a one-dimensional curve. One of the two-dimensions of the surface seems to be spontaneously compactified, and the remaining dimension survives along the curve. A spontaneous breakdown of the rotational symmetry is also found in the tubular phase, where the surface spans a tubular surface along a one-dimensional straight line. 

Moreover, the phase transition is reflected not only in the surface shape but also in the surface structure. In fact, the histogram $h(L)$ of the bond length varies against $\alpha$ and, $h(L)$ in one phase is different from those in the other phases, although $h(L)$ seems smoothly varies even at the transition point between the linear phase and the tubular phase. This point is in sharp contrast to that of the fixed connectivity model, where $h(L)$ changes discontinuously at the transition point between the tubular phase and the smooth phase \cite{KOIB-PRE-2004-2}. On the contrary, the histogram $h(A)$ of the area $A$ is not influenced by phase transitions as in the fixed connectivity model, and no difference can be seen in $h(A)$ in four different phases.

This work is supported in part by a Grant-in-Aid for Scientific Research from Japan Society for the Promotion of Science.  




\begin{thebibliography}{999}
\bibitem{WHEATER-JP1994}
 J.F. Wheater, J. Phys. A Math. Gen. {\bf 27}, 3323 (1994).

\bibitem{NELSON-SMMS2004-1}
D. Nelson, in {Statistical Mechanics of Membranes and Surfaces, Second Edition}, edited by  D. Nelson, T. Piran, and S. Weinberg, p.1 (World Scientific, 2004). 

\bibitem{David-TDQGRS-1989}
F. David,  in {Two dimensional quantum gravity and random surfaces, Vol.8}, edited by  D. Nelson, T. Piran, and S. Weinberg, p.81 (World Scientific, Singapore, 1989).

\bibitem{NELSON-SMMS2004-149}
D. Nelson, in {Statistical Mechanics of Membranes and Surfaces, Second Edition}, edited by  D. Nelson, T. Piran, and S. Weinberg, p.149 (World Scientific, 2004). 

\bibitem{Wiese-PTCP2000}
K. Wiese, in: C. Domb, J.L. Lebowitz  (Eds.), Phase Transitions and Critical Phenomena, Vol. 19, p.253 (Academic Press, London, 2000).

\bibitem{Bowick-PREP2001}
 M. Bowick and A. Travesset, Phys. Rep. {\bf 344}, 255 (2001).

\bibitem{Gompper-Schick-PTC-1994}
G. Gompper and M. Schick, \textit{Self-assembling amphiphilic systems}, In
\textit{Phase Transitions and Critical Phenomena 16}, edited by C. Domb and J.L. Lebowitz, p.1 (Academic Press, 1994).

\bibitem{HELFRICH-1973}
 W. Helfrich, Z. Naturforsch, {\bf 28c}, 693 (1973).

\bibitem{POLYAKOV-NPB1986}
 A.M. Polyakov, Nucl. Phys. B {\bf 268}, 406 (1986).

\bibitem{KLEINERT-PLB1986}
 H. Kleinert, Phys. Lett. {\bf 174B}, 335 (1986).

\bibitem{Peliti-Leibler-PRL1985}
 L. Peliti and S. Leibler, Phys. Rev. Lett. {\bf 54} (15), 1690 (1985).

\bibitem{DavidGuitter-EPL1988}
 F. David and E. Guitter, Europhys. Lett,  {\bf 5} (8), 709 (1988).

\bibitem{PKN-PRL1988}
M. Paczuski, M. Kardar, and D. R. Nelson, Phys. Rev. Lett. {\bf 60}, 2638 (1988).

\bibitem{BKS-PLA2000}
 M.E.S. Borelli, H. Kleinert, and Adriaan M.J. Schakel, Phys. Lett. A {\bf 267}, 201 (2000).

\bibitem{BK-PRB2001}
 M.E.S. Borelli and H. Kleinert, Phys. Rev. B {\bf 63}, 205414 (2001). 

\bibitem{GREST-JPIF1991}
G. Grest, J. Phys. I (France) {\bf 1}, 1695 (1991).

\bibitem{BOWICK-TRAVESSET-EPJE2001}
M. Bowick and A. Travesset,  Eur. Phys. J. E {\bf 5}, 149 (2001).

\bibitem{BCTT-PRL2001}
M. Bowick, A. Cacciuto, G. Thorleifsson, and  A. Travesset, Phys. Rev. Lett. {\bf 87}, 148103 (2001).

\bibitem{KANTOR-NELSON-PRA1987}
 Y. Kantor and  D.R. Nelson, Phys. Rev. A {\bf 36}, 4020 (1987).

\bibitem{KANTOR-SMMS2004}
Y. Kantor, in {Statistical Mechanics of Membranes and Surfaces, Second Edition}, edited by  D. Nelson, T. Piran, and S. Weinberg, p.111 (World Scientific, 2004). 

\bibitem{WHEATER-NPB1996}
J.F. Wheater, Nucl. Phys. B {\bf 458}, 671 (1996).

\bibitem{BCFTA-JP96-NPB9697}
 M. Bowick,  S. Catterall,  M. Falcioni,  G. Thorleifsson, and K. Anagnostopoulos, J. Phys. I France  {\bf 6}, 1321 (1996);\\
 M. Bowick,  S. Catterall,  M. Falcioni,  G. Thorleifsson, and K. Anagnostopoulos, Nucl. Phys. Proc. Suppl. {\bf 47}, 838 (1996);\\ 
 M. Bowick,  S. Catterall,  M. Falcioni,  G. Thorleifsson, and  K. Anagnostopoulos, Nucl. Phys. Proc. Suppl. {\bf 53}, 746 (1997).

\bibitem{CATTERALL-NPBSUP1991}
 S.M. Catterall, J.B. Kogut, and R.L. Renken, Nucl. Phys. Proc. Suppl. B {\bf 99A}, 1 (1991).

\bibitem{AMBJORN-NPB1993}
 J. Ambjorn, A. Irback, J. Jurkiewicz, and B. Petersson, Nucl. Phys. B {\bf 393}, 571 (1993).

\bibitem{ABGFHHM-PLB1993}
 K. Anagnostopoulos, M. Bowick, P. Gottington, M. Falcioni, L. Han, G. Harris, and E. Marinari, Phys. Lett. {\bf 317B}, 102 (1993).

\bibitem{BCHHM-NPB9393}
 M. Bowick, P. Coddington, L. Han, G. Harris, and E. Marinari, Nucl. Phys. Proc. Suppl. {\bf 30}, 795 (1993);\\ \noindent
 M. Bowick, P. Coddington, L. Han, G. Harris, and  E. Marinari, Nucl. Phys. B {\bf 394}, 791 (1993).

\bibitem{KOIB-PLA-20023}
 H. Koibuchi,  Phys. Lett. A {\bf 300}, 582 (2002); \\
 H. Koibuchi, N. Kusano, A. Nidaira, K. Suzuki, and M.Yamada,  Phys. Lett. A {\bf 319}, 44 (2003).

\bibitem{KOIB-PLA-2004}
 H. Koibuchi, N. Kusano, A. Nidaira, and K. Suzuki,  Phys. Lett. A {\bf 332}, 141 (2004).

\bibitem{KOIB-EPJB-2005}
 H. Koibuchi, Eur. Phys. J. B \textbf{45}, 377 (2005).

\bibitem{KOIB-EPJB-2006}
  H. Koibuchi,  Eur. Phys. J. B \textbf{52}, 265 (2006).
\bibitem{KD-PRE2002}
J-P. Kownacki and H. T. Diep, Phys. Rev. E {\bf 66}, 066105 (2002).

\bibitem{KOIB-PRE-2004-1}
H. Koibuchi, N. Kusano, A. Nidaira, K. Suzuki, and M. Yamada, Phys. Rev. E {\bf 69}, 066139
(2004).

\bibitem{KOIB-PRE-2005-1}
 H. Koibuchi and T. Kuwahata, Phys. Rev. E {\bf 72}, 026124 (2005). 

\bibitem{KOIB-NPB-2006}
 I. Endo and H. Koibuchi,  Nucl. Phys. B {\bf 732 [FS]}, 426 (2006).

\bibitem{KOIB-PRE-2007-1}
 H. Koibuchi,  Phys. Rev. E {\bf 75}, 011129 (2007).

\bibitem{KOIB-PRE-2003}
 H. Koibuchi, A. Nidaira, T. Morita, and K. Suzuki,
Phys. Rev. E  \textbf{68}, 011804 (2003). 

\bibitem{KOIB-PRE-2004-2}
 H. Koibuchi,  Z. Sasaki, and K. Shinohara, 
Phys. Rev. E  \textbf{70}, 066144 (2004). 

\bibitem{BJ-PRD-1993-1994}
C.F. Baillie, and D.A. Johnston,  Phys. Rev. D {\bf 48}, 5025 (1993); {\bf 49}, 4139 (1994).

\bibitem{BEJ-PLB-1993}
C.F. Baillie, D. Espriu, and D.A. Johnston, Phys. Lett. {\bf 305B}, 109 (1993).

\bibitem{BIJJ-PLB-1994}
C.F. Baillie, A. Irback, W. Janke and D.A. Johnston, Phys. Lett. {\bf 325B}, 45 (1994).

\bibitem{FW-PLB-1993}
N. Ferguson and J.F. Wheater, Phys. Lett. {\bf 319B}, 104 (1993).

\bibitem{Nambu}
Y. Nambu, lectures at Copenhagen Symposium (1970), in  Broken Symmetry, Selected Papers of Y Nambu, World Scientific Series in 20th Century Physics Vol. 13, edited by T. Eguchi and K. Nishijima, p.280 (World Scientific, 1995); 
T. Goto, Prog. Theor. Phys. {\bf 46}, 1560 (1971).

\bibitem{ADF}
J. Ambjorn, B. Durhuus and J. Frohlich, Nucl. Phys. B {\bf 257}, 433 (1985).

\bibitem{Polyakov-contempV3-1987}
A. Polyakov, Gauge Fields and Strings, Contemporary Concepts in Physics Vol.3, p.254 (Harwood Academic Publishers, 1987).

\bibitem{David-NP}
F. David, Nucl. Phys. B {\bf 257} [FS14] 543 (1985).

\bibitem{BKKM}
 D.V. Boulatov, V.A. Kazakov, I.K. Kostov and A.A. Migdal, Nucl. Phys. B {\bf 275 [FS17]},  641 (1986).

\bibitem{Matsumoto-Nishimura-1998}
M. Matsumoto and T. Nishimura, "Mersenne Twister: A 623-dimensionally equidistributed uniform pseudorandom number generator", ACM Trans. on Modeling and Computer Simulation Vol. 8, No. 1, pp.3-30, January (1998).

\end{thebibliography}
\end{document}